\documentclass[sigconf]{acmart}
\AtBeginDocument{%
  }

\usepackage{tikz}
\usetikzlibrary{positioning,arrows.meta,shapes.geometric,fit,backgrounds}
\usepackage{booktabs}
\usepackage{graphicx}

\copyrightyear{2026}
\acmYear{2026}
\setcopyright{cc}
\setcctype{by-nc-nd}
\acmConference[CIKM '26]{Proceedings of the 35th ACM International Conference on Information and Knowledge Management}{November 07--11, 2026}{Rome, Italy}
\acmBooktitle{Proceedings of the 35th ACM International Conference on Information and Knowledge Management (CIKM '26), November 07--11, 2026, Rome, Italy}
\acmDOI{10.1145/3799682.3840239}
\acmISBN{979-8-4007-2539-5/2026/11}

\begin{document}

%% ============================================================
\title{EviMap: Evidence-Grounded Hierarchical Topic Maps for Exploring Unlabeled Corpora}

%% SINGLE-BLIND: put real authors here
\author{Zhiyin Tan}
\correspondingauthor
\orcid{0009-0002-4166-5810}
\affiliation{%
  \institution{L3S Research Center, Leibniz University Hannover}
  \city{Hannover}
  \country{Germany}}
\email{zhiyin.tan@l3s.de}

\author{Changxu Duan}
\orcid{0000-0003-0547-0901}
\correspondingauthor
\affiliation{%
  \institution{Technische Universität Darmstadt}
  \city{Darmstadt}
  \country{Germany}}
\email{duan@linglit.tu-darmstadt.de}

\renewcommand{\shortauthors}{Zhiyin Tan and Changxu Duan}

%% ============================================================
\begin{abstract}
Research teams and organizations often explore unfamiliar free-text collections, from survey comments and reviews to reports and domain documents, before labels, queries or coding schemes exist. At this stage, the first thematic map shapes what users notice, prioritize and carry into downstream analysis, so it should be trusted only insofar as it can be verified. Existing options force a trade-off between scale and verifiability. Qualitative coding preserves evidence but is slow. Search presupposes a query. Clustering and topic models scale but produce labels users must interpret. One-shot large language model (LLM) summaries are fluent yet difficult to reproduce or audit.
We present EviMap, an interactive system providing researchers and practitioners with an auditable thematic overview of such corpora. Guided by model-generated context describing the corpus and hypothesized stakeholder concerns, EviMap extracts within-document evidence phrases and organizes them, rather than whole documents, into a three-level map of aspects, groups and fine-grained topics. Embedding-based clustering narrows the search space for finer semantic judgments by the LLM. Each node traces back to supporting phrase spans, so documents link to topics through evidence they contain and users can audit labels against the original text.
Users can start from a top-level corpus map, drill into topics, inspect highlighted evidence in original documents, and combine two topics to find documents discussing both. We demonstrate this workflow across six heterogeneous corpora spanning 2,108 to 101,699 documents, with a comparison against flat and hierarchical LLM baselines. By grounding every label in verbatim source spans, EviMap makes a topic map not just readable, but verifiable.
Code, demo video, and interactive dashboard are available at \url{https://github.com/zhiyintan/EviMap}.

\end{abstract}

%% CCS



\maketitle

%% ============================================================
\section{Introduction}

Large free-text collections often contain what structured fields or pre-existing taxonomies miss, such as unexpected concerns, explanations, edge cases, and strong minority viewpoints.
Open-ended survey responses, interview transcripts, and first-person narratives surface issues that closed questions cannot~\cite{ocathain2004,chandrasekar2024}.
At scale, exhaustive reading becomes impractical. One health-services project faced over half a million free-text comments across 320,500 surveys~\cite{etz2018}.
These collections pose a joint scale and vocabulary problem. Analysts must discover concepts before knowing which labels, queries, or codes are worth using.
A weak first map can hide minority concerns or anchor later coding to accidental labels.

This paper targets early corpus understanding, the task of making sense of an unlabeled text collection at the start of a project, before a taxonomy, coding scheme, or precise query exists.
They want to know which themes recur, how those themes relate, which issues deserve closer inspection, and which passages support each theme.
A useful system must therefore provide an overview without hiding the passages that make the overview credible.
This is the exploratory search setting that Marchionini distinguishes from known-item lookup~\cite{marchionini2006}, and it corresponds to the schema-building stage of the sensemaking process~\cite{klein2006}.
The audience is researchers and practitioners in computational social science, digital humanities, market research, journalism, and intelligence analysis.

\begin{table}[t]
\centering
\caption{Positioning against nearby LLM-based systems using demo-facing criteria.
Limited means the capability is possible, but not the primary artifact or interaction model.}
\label{tab:positioning}
\small
\setlength{\tabcolsep}{2pt}
\begin{tabular}{lcccc}
\toprule
System & Artifact & Map & Support & Stabilizer \\
\midrule
TopicGPT~\cite{pham2024} & topics & limited & quotes & none \\
TnT-LLM~\cite{wan2024} & taxonomy & no & no spans & iterative \\
LLooM~\cite{lam2024} & concepts & concept UI & examples & none \\
Chain-of-Layer~\cite{zeng2024} & taxonomy & no & no spans & ensemble \\
EviMap & dashboard & 3-level & spans & co-assoc. voting \\
\bottomrule
\end{tabular}
\vspace{-4mm}
\end{table}

This scale-verifiability trade-off remains unresolved.
Manual thematic analysis and open coding preserve context~\cite{braun2006}, but do not scale to the volumes at which teams now collect free text.
Search returns ranked documents and presupposes a query, so it organizes nothing for a user who cannot yet articulate one~\cite{marchionini2006,whiteroth2009}.
Document clustering produces partitions, but shifts the labeling problem to the user.
Topic models, from LDA to neural variants such as the embedded topic model~\cite{dieng2020} and BERTopic~\cite{grootendorst2022}, emit bag-of-words topics that require human interpretation~\cite{chang2009}, are unstable across runs, and whose automated coherence scores correlate weakly with human judgment for modern models~\cite{hoyle2021}.
They also provide no native link from a topic to the exact phrase that motivated it.
More fundamentally, document-level models force each document under ranked topics, so topics match loosely, absorb noise, wash out distinctive details, and may include documents that never raised them.
One-shot LLM summaries are readable but unstructured, non-reproducible, prone to hallucination~\cite{ji2023,maynez2020}, and difficult to verify against the source.
Recent evidence that LLM coding errors can introduce biases in qualitative analysis argues against treating generated labels as uninspected ground truth~\cite{ashwin2025}.

EviMap closes this gap by joining two lines of work, LLM-based topic induction with readable labels~\cite{pham2024,wan2024,zhang2023,wang2023goalex,lam2024,zeng2024} and attributed generation grounded in source spans~\cite{rashkin2023,bohnet2022,gao2023,slobodkin2024}.
The result is a persistent, browsable, multi-level topic map whose leaf topics are grounded in highlighted phrase spans and whose higher-level groups inherit evidence from their descendants.
Crucially, EviMap groups evidence phrases rather than whole documents, so every document under a node genuinely raised it, preventing spurious associations.
A document can also appear under many nodes, keeping its heterogeneous detail in the map instead of collapsing it into one label.
The demo treats every generated label as a claim the user can check. Attendees move from an automatically induced corpus map to the phrases and documents behind each label, test topic combinations as exploratory queries, and examine unresolved long-tail material.

This paper makes three contributions:

\textbf{(1) An evidence-grounded map design.} The hierarchical map keeps every node traceable to verbatim source spans and links documents to topics only through evidence they contain.

\textbf{(2) A phrase-level hierarchy induction method.} EviMap builds the hierarchy bottom-up from evidence phrases by combining embedding-based clustering with LLM semantic judgments, without imposing a predefined taxonomy.

\textbf{(3) An auditable exploration system.} The dashboard supports map browsing, evidence inspection, topic-intersection retrieval, and long-tail auditing across six heterogeneous corpora.

%% ============================================================
\section{Related Work}

\noindent\textbf{Exploratory sensemaking and qualitative tools.}
Exploratory search, sensemaking, and qualitative analysis frame how users organize a corpus before formulating precise queries~\cite{marchionini2006,bates1989,whiteroth2009,klein2006}.
Manual thematic analysis and open coding remain standard~\cite{braun2006}, and tools such as NVivo, MAXQDA, ATLAS.ti, and Dovetail help organize codes and excerpts but still require user-curated coding schemes.
Interactive systems research explores LLM-assisted qualitative annotation~\cite{xiao2023,dai2023,gebreegziabher2023}.
EviMap targets earlier discovery by inducing a candidate map grounded in inspectable evidence.

\noindent\textbf{Topic models and visual analytics.}
Classical and neural topic models~\cite{dieng2020,grootendorst2022} produce word-distribution topics with limitations in interpretability~\cite{chang2009}, evaluation~\cite{hoyle2021}, and stability.
Visual analytics systems such as Termite~\cite{chuang2012}, Serendip~\cite{alexander2014}, and TopicLens~\cite{kim2017} inspect topic-model outputs.
EviMap instead clusters LLM-extracted evidence phrases, making phrases the clustering unit rather than documents and the attribution unit rather than topic words.

\noindent\textbf{LLM-based topic and taxonomy systems.}
TopicGPT~\cite{pham2024} , the closest prior system, generates natural-language topics with supporting quotations and can be extended hierarchically.
EviMap instead makes a browsable three-level map the primary artifact, grounds leaf topics in highlighted sub-sentence spans, and aggregates repeated grouping decisions through co-association voting.
TnT-LLM combines taxonomy generation with classifier construction~\cite{wan2024}.
ClusterLLM and GoalEx guide or explain clustering~\cite{zhang2023,wang2023goalex}.
LLooM induces high-level concepts~\cite{lam2024}.
Chain-of-Layer builds taxonomies layer by layer with an ensemble filter~\cite{zeng2024}.

Table~\ref{tab:positioning} compares the closest systems using demo-facing criteria rather than a full capability ranking.
Map denotes a persistent topic map, Support names the evidence exposed to users, and Stabilizer names an explicit mechanism against unstable grouping.

%% ============================================================
\section{System Architecture}

\begin{figure}[t]
\centering
\includegraphics[width=0.98\linewidth]{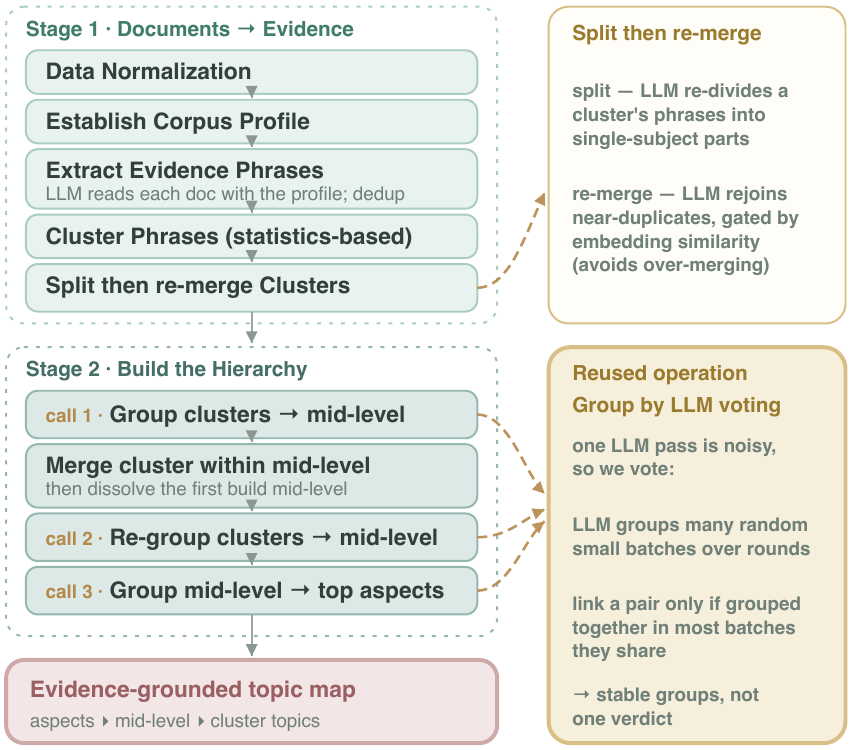}
\vspace{-3mm}
\caption{The EviMap pipeline. Stage 1 extracts evidence phrases and clusters them. Stage 2 builds the aspect hierarchy by repeatedly calling one LLM grouping operation (right). Categories grow bottom-up from evidence.}
\vspace{-4mm}
\label{fig:pipeline}
\end{figure}

EviMap turns an unlabeled corpus into an evidence-grounded topic map, a browsable hierarchy whose nodes trace back to evidence phrases and document spans.
The map is induced in two stages (Figure~\ref{fig:pipeline}), evidence-phrase extraction followed by bottom-up structure discovery from that evidence.

\begin{figure*}[t]
\centering
\includegraphics[width=0.98\textwidth]{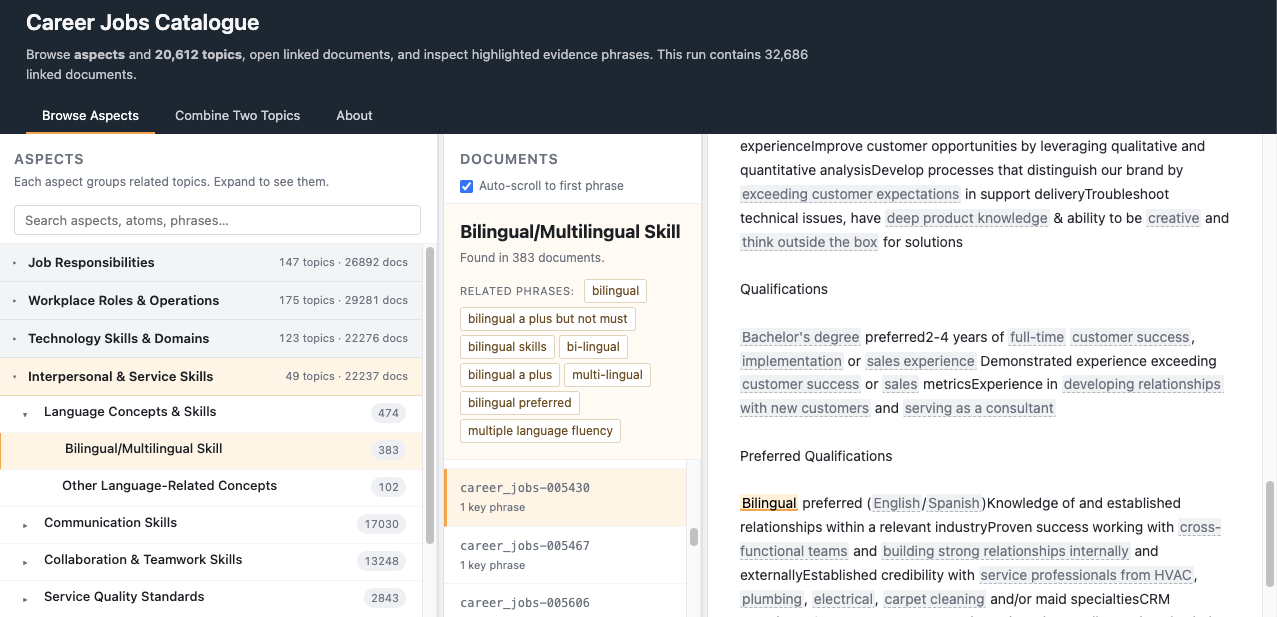}
\vspace{-2mm}
\caption{Career-jobs dashboard with Bilingual/Multilingual Skill selected under Interpersonal \& Service Skills and \texttt{career\_jobs-005430} open.
Current-topic evidence is orange, other-topic phrases gray and clickable.}
\label{fig:dashboard}
\vspace{-2mm}
\end{figure*}

\noindent\textbf{(1) Data normalization}.
Heterogeneous corpora are normalized to a common record format
$\{$\texttt{doc\_id}, \texttt{text}, \texttt{metadata}$\}$.
\noindent\textbf{(2) Evidence-phrase extraction}.
From a small sample, the LLM builds a lightweight domain profile and hypothesizes which aspects different stakeholder roles might prioritize, then uses both to guide the extraction of short evidence phrases, the atomic unit of evidence, from each document. These model-generated inputs guide extraction only. The final categories are induced from extracted evidence, not from a preset list.
\noindent\textbf{(3) Phrase index and embeddings}.
Extracted phrases are expanded into a phrase~$\times$~document occurrence table.
We deduplicate, compute IDF, record the character offset of each phrase in each document, and embed every phrase.
The stored offsets turn each occurrence into a span that can be highlighted in its original context.
\noindent\textbf{(4) Bottom-up grouping}.
The core of the second stage is a reusable operation that turns a flat set of items into semantic groups.
K-Means first forms deliberately fine-grained embedding clusters.
These are not final groups. They shrink the search space so each LLM call sees a small, relevant context instead of the whole corpus.
Within that context, the LLM makes the finer semantic judgments embeddings cannot.
This design bounds the damage of any single LLM decision, because a poor coarse cluster is more likely to leave related items separated than to trigger corpus-wide merge errors.
Grouping is stabilized by multi-round voting.
Rather than organizing one long list in one pass, EviMap samples small random subsets and asks the LLM to group each subset.
For every pair, it computes a co-association score from grouped-together counts divided by co-sampled counts.
Only pairs above the threshold are linked, resisting over-aggressive merging in which a few spurious links pull unrelated items into one group.
\noindent\textbf{(5) Fine-grained topics}.
We apply this operation to the evidence phrases to obtain fine-grained topics.
Near-duplicates are merged, and very small topics are folded into related topics or retained in a visible Other bin for inspection.
Each surviving topic carries a label, its supporting documents, representative evidence phrases, and the phrase spans that ground it in the text.
\noindent\textbf{(6) Hierarchy}.
EviMap builds the map in two hierarchy passes.
It first groups cleaned fine-grained topics into mid-level groups, then clusters those groups into a small fixed number of top-level aspects.
Reusing one operation keeps the passes consistent, and fixing the aspect count prevents a single overly broad cluster.
All naming and grouping prompts state only abstract rules with no in-domain examples, so the procedure transfers across corpora.
Labels are encouraged to name the concrete object, action, or relation being discussed rather than generic headings such as Miscellaneous Concepts.

%% ============================================================
\section{The Interactive Dashboard}

The induced map serves as an exploration dashboard (Figure~\ref{fig:dashboard}).
The dashboard is not an editor. Users inspect and challenge the candidate map without committing merge, split, or rename operations.
The left pane expands aspects into groups and fine-grained topics, the topic card shows support counts and representative phrases, and the document pane highlights evidence phrases in context.

The dashboard also supports compositional exploration.
Users can combine two topics to retrieve documents that discuss both, such as job postings mentioning both bilingual skills and customer-facing work.
The Other bin, representative phrases, support counts, and highlighted evidence help users audit uncertain labels.
Together, these interactions keep generated labels as auditable claims rather than labels to accept blindly.
Curation operations such as rename, merge, split, and export are deliberately out of scope for this demo and planned as future work.

\section{Demonstration Scenario}

The booth demonstration begins with the job-posting dashboard as job postings provide a broadly shared context, allowing attendees to assess directly.
We open the view in Figure~\ref{fig:dashboard}, then drill from Interpersonal \& Service Skills through Language Concepts \& Skills to Bilingual/Multilingual Skill.
The topic card reports 383 supporting documents and representative phrases including bilingual, bilingual skills, bi-lingual, bilingual a plus, and multi-lingual.
The attendee then clicks the document \texttt{career\_jobs-005430}.
The right pane highlights current-topic evidence in orange and other-topic evidence in gray, including customer success and strong organizational skills.
Clicking a gray span jumps to that topic.

Next, the attendee uses the Combine Two Topics tab.
For example, selecting Bilingual/Multilingual Skill and Customer Success with a 200-character window returns job postings that discuss both language requirements and customer-facing work.
The same document pane then highlights evidence for both topics, which turns a broad map into a composable corpus search tool.
We then switch to the peer-review dashboard and drill from Mathematical Foundations to Distribution Types and Roles to show that the prompts and grouping rules are not job-specific.
Finally, we expose the Other bin and the run summary so attendees can see unresolved long-tail topics rather than hiding them.

%% ============================================================
\begin{table}[t]
\centering
\caption{Corpora used in the submitted artifact.
Docs counts documents with at least one extracted evidence phrase.}
\label{tab:corpora}
\small
\setlength{\tabcolsep}{3pt}
\begin{tabular}{@{}p{0.45\columnwidth}p{0.28\columnwidth}r@{}}
\toprule
Corpus & Domain & Number of docs \\
\midrule
ICLR peer reviews~\cite{datasetIclrPeerReviews} & Scholarly reviews & 2,108 \\
TripAdvisor Hotel reviews~\cite{datasetTripadvisorReviewRating} & Hospitality reviews & 2,864 \\
Pitchfork music reviews~\cite{datasetPitchforkReviews} & Music criticism & 24,640 \\
LinkedIn job postings~\cite{datasetLinkedinJobPostings} & Labor-market text & 32,686 \\
Recipes~\cite{datasetRecipesNutrition} & Instructional text & 39,447 \\
EU debates~\cite{datasetEuDebates} & Political speech & 101,699 \\
\bottomrule
\end{tabular}
\vspace{-5mm}
\end{table}
%% ============================================================

%% ============================================================
\section{Demonstration Corpora}

Table~\ref{tab:corpora} summarizes coverage for six corpora spanning heterogeneous domains, all processed with the same evidence-grounded workflow. 

\noindent \textbf{Artifact traceability validation.}
Purpose-oriented evaluation judges topics by their intended use, not a universal score~\cite{tan2025purpose, tan2025bridging}, so we evaluate EviMap for auditable exploration.
Following work on attributed and verifiable generation, we treat a topic label as auditable only if the dashboard resolves it to source spans in the original documents~\cite{rashkin2023,bohnet2022,gao2023,slobodkin2024}.
The validation covered 3,290,920 topic-evidence links and 4,357,621 character-offset spans across six dashboards.
It found no missing document references or span-text mismatches at the stored offsets.
These checks validate traceability, not topic quality.

\noindent \textbf{Positioning against LLM baselines.}
To compare artifact shape, we ran both baselines on each dashboard's full document cohort using the same LLM backend.
On job postings, TopicGPT-style yields 24 flat topics and TnT-LLM-style 8 top-level and 32 child categories, versus EviMap's 14 induced aspects plus one Other bin, 350 groups, and 3,245 leaf nodes.
Following this view, we adapted an LLM-as-judge diagnostic~\cite{tan2025purpose} to score each topic unit on a 1--5 scale for evidence coherence and label soundness.
Across the six corpora, 96 stratified non-Other EviMap leaf topics per corpus averaged 4.53/4.33 (4.30/4.04 with Other bins), versus 3.60/3.50 for TopicGPT-style topics and 3.96/3.79 for TnT-LLM-style child categories.
EviMap leads on coherence in five corpora and on label soundness in four, with TopicGPT-style ahead on hotel reviews.
The diagnostic characterizes semantic soundness, not topic accuracy.

%% ============================================================
\section{System Realization}

EviMap is implemented as a reproducible pipeline from normalized document records to the materialized dashboard.
LLM calls use deterministic decoding through a provider-agnostic API.
The submitted builds use \texttt{DeepSeek-V4-Flash}~\cite{deepseekai2026deepseekv4}, chosen for its low per-call cost over the many short prompts the pipeline issues, and the local multilingual encoder \texttt{paraphrase-\allowbreak multilingual-\allowbreak MiniLM-\allowbreak L12-v2}~\cite{reimers2019sentencebert}, so phrase similarity transfers across languages. Both are swappable.

The submitted dashboards use five voting rounds and a minimum co-association threshold of 0.7 for the first hierarchy layer.
The top layer uses fixed-$K$ agglomerative clustering over voted similarities.
We use three levels and constrain $K \leq 20$ as defaults chosen through preliminary exploration to balance thematic coverage against navigational complexity~\cite{jacko1996hierarchical}.
Explicit Other bins are displayed separately when present.
For the largest corpus (101,699 documents), a complete build makes ${\sim}1.1\times10^5$ LLM calls over ${\sim}4.5\times10^8$ tokens in about 18 hours on 4 $\times$ H100 GPUs with 128 concurrent extraction workers.
Extraction makes one call per document and dominates the workload, so runtime is approximately linear in document count.
Run scripts, model settings, prompts, intermediate files, and provenance logs are part of the submitted artifact.

\section{Conclusion}

%EviMap addresses the stage before analysis has stable labels, queries, or a coding scheme. It turns an unlabeled corpus into a three-level topic map whose leaf nodes are grounded in evidence-phrase spans and whose higher-level nodes remain auditable through descendant evidence. Across six corpora, the artifact presents generated labels as auditable claims rather than a gold taxonomy. Current limitations include reliance on LLM semantic judgments, which co-association voting mitigates but cannot eliminate, and the lack of standard gold hierarchies for automatic evaluation. Semi-structured corpora with existing sectioning could serve as proxy ground truth. Beyond the curation features of Section 4, future work includes a formative user study, sensitivity and run-to-run variance analysis of the voting parameters, and systematic evaluation on multilingual corpora.

EviMap addresses exploratory analysis before labels, queries, or coding schemes are established.
It builds a three-level map bottom-up from evidence phrases, keeping every node traceable to source spans and linking documents to topics only through evidence they contain.
Across six corpora, the dashboard supports map browsing, evidence inspection, topic intersection, and long-tail auditing while presenting generated labels as auditable claims rather than a gold taxonomy.
Several limitations remain. Model-generated extraction context may omit important aspects, small-topic folding may obscure minority concerns, and LLM grouping and labeling remain fallible.
Gold hierarchies for automatic evaluation are also missing, though sectioned corpora may provide partial proxies.
Future work will therefore add user-specified aspects and reintegration of Other material, measure extraction recall, parameter sensitivity, and run-to-run variation, and broaden validation with formative user studies and multilingual evaluation.
%Together these steps aim to make the first thematic map of an unfamiliar corpus scalable and verifiable at once.

\begin{acks}
Zhiyin Tan was funded by the ``HybrInt - Hybrid Intelligence through Interpretable AI in Machine Perception and Interaction'' project (Zukunft Nds, Niedersächsisches Ministerium für Wissenschaft, Grant ID: ZN4219), co-funded by the German Research Foundation (DFG) - project NFDI4DataScience (No. 460234259).
Changxu Duan was funded by the Federal Ministry of Education and Research (BMBF) under grant no. 01UG2130A. 
We gratefully acknowledge support from the hessian.AI Service Center (funded by the Federal Ministry of Research, Technology and Space, BMFTR, grant no. 16IS22091) and the hessian.AI Innovation Lab (funded by the Hessian Ministry for Digital Strategy and Innovation, grant no. S-DIW04/0013/003), and computing time granted by Emmy/Grete at NHR-Nord@Göttingen (project \verb|nhr_he_starter_25563|).
\end{acks}
%% ============================================================
%% GenAI Usage Disclosure (UNLIMITED length, does not count toward 4 pages)
%% Edit to reflect your actual use.
%% Note: the LLM components that are part of the SYSTEM are described in the body above.
%% This section concerns the authors' use of GenAI tools while preparing the paper and research.
\section*{GenAI Usage Disclosure}
The EviMap system itself uses large language models as core technical components (phrase extraction, semantic cleanup, naming, and multi-round voting), as described in Sections~3 to~6.
In preparing this manuscript, the authors used a generative AI assistant for grammar and style improvements.
All research ideas, system design, experiments, claims, and final text were conceived, verified, and approved by the authors, who take full responsibility for the content, including the accuracy of all references.

%% ============================================================
%% References (UNLIMITED, do not count toward 4 pages)
%% Entries live in the shared references.bib.
%% Verify each against DBLP / the original before submission.
\bibliographystyle{ACM-Reference-Format}
\bibliography{references}

\end{document}